\documentclass[
 reprint,
 amsmath,
 pre,
 amssymb,
 nofootinbib,
 aps,prd,twocolumn]{revtex4-1}

\usepackage[pdftex]{graphicx}
\usepackage{wasysym}
\usepackage{amsfonts}
\usepackage{ifsym}
\usepackage{hyperref} 
\usepackage{slashed}

\usepackage[utf8]{inputenc}
\usepackage[bb=boondox]{mathalfa}

\def\d{\partial}
\def\l{\left(}
\def\r{\right)}

\newcommand{\be}{\begin{equation}}
\newcommand{\ee}{\end{equation}}
\newcommand{\bea}{\begin{eqnarray}}
\newcommand{\eea}{\end{eqnarray}}
\newcommand{\bg}{\begin{gather}}
\newcommand{\eg}{\end{gather}}
\newcommand{\bseq}{\begin{subequations}}
\newcommand{\eseq}{\end{subequations}}

\begin{document}
\title{
 Classical Integrability of the Zigzag Model
 }
\author{John C. Donahue}
\author{ 
Sergei Dubovsky
}
\affiliation{
  Center for Cosmology and Particle Physics, Department of Physics,
      New York University,
      New York, NY, 10003, USA
      }

\begin{abstract}
The zigzag model is a relativistic $N$-body system arising in the high energy limit  of the worldsheet scattering in adjoint two-dimensional
QCD. We prove classical Liouville integrability of this model by providing an explicit construction of $N$ charges in involution. Furthermore, we also prove that the system is maximally
superintegrable by constructing $N-1$ additional independent charges. All of these charges are piecewise linear functions of coordinates and momenta. The classical time delays are determined algebraically from this integrable structure.
The resulting $S$-matrix is the same as in the $N$-particle subsector of a massless $T\bar{T}$ deformed fermion.
\end{abstract}
\maketitle
\section{Introduction}
Understanding the mechanism of quark confinement continues to stand out as an interesting challenge and as a source of new developments in theoretical and mathematical physics.

Starting with \cite{Dubovsky:2012sh}, much effort has recently been put  into a study of scattering on the worldsheet of a single confining string in the t' Hooft planar limit  \cite{'tHooft:1973jz}. An especially  interesting problem is to understand the high energy dynamics on the worldsheet. In this regime one expects the confining string to exhibit  characteristically gravitational behavior similarly to that of
critical strings \cite{Dubovsky:2012wk}.

An interesting possibility is that hard high energy scattering on the worldsheet approaches integrable asymptotics \cite{Dubovsky:2015zey}. A concrete version of this proposal, the Axionic String Ansatz (ASA)\cite{Dubovsky:2015zey,Dubovsky:2016cog}, identifies the corresponding integrable theories as $T\bar{T}$-deformations \cite{Zamolodchikov:2004ce,Dubovsky:2013ira,Smirnov:2016lqw,Cavaglia:2016oda} of certain free massless models.
This proposal is motivated and supported by the recent analysis \cite{Dubovsky:2013gi,Dubovsky:2014fma,Dubovsky:2015zey} of lattice measurements
of flux tubes excitations \cite{Athenodorou:2010cs,Athenodorou:2011rx,Athenodorou:2016kpd,Athenodorou:2017cmw}. For $D=3$ gluodynamics it is also supported by lattice determinations of glueball masses and spins \cite{Athenodorou:2016ebg,Dubovsky:2016cog,Conkey:2019blu}. The physical reason for the emergence of integrable dynamics in the high energy worldsheet scattering is the asymptotic freedom of the underlying gauge theory \cite{Dubovsky:2018vde}. Finally, axionic strings also came out recently as a result of the flux tube $S$-matrix bootstrap \cite{EliasMiro:2019kyf}.

A natural playground for testing this idea is provided by adjoint QCD in $D=2$ dimensions ($aQCD_2$). The spectrum of this theory has been extensively studied  in early 90's \cite{Dalley:1992yy,Bhanot:1993xp,Kutasov:1993gq,Kutasov:1994xq} (see, e.g.,  \cite{Katz:2013qua,Cherman:2019hbq} for more recent interesting works). A study of the worldsheet scattering in the model has been initiated  in \cite{Dubovsky:2018dlk}, building up on the techniques developed in prehistoric times \cite{Coleman:1976uz,Witten:1978ka}. 

Recently,  a candidate relativistic $N$-body system describing the integrable high-energy  asymptotics of the worldsheet theory in $aQCD_2$ has been identified \cite{Donahue:2019adv}. For reasons which will become clear we  call this system the ``zigzag model". In \cite{Donahue:2019adv} we provided partial numerical and analytical evidence for classical integrability of the zigzag model. The goal of the present paper is to provide a complete proof that the zigzag model is integrable at the classical level.

The rest of the paper is organized as follows. In section~\ref{sec:model} we describe the model.
In section~\ref{sec:top} we describe a discrete topological invariant present in the model. This topological invariant ensures that the total number of left- and right-movers is conserved in any scattering event.
In section~\ref{sec:linear} we construct $N$ conserved charges in involution and  $2N-1$ algebraically independent conserved charges. All of these charges are piecewise linear functions of coordinates and momenta. This construction establishes that the zigzag model is Liouville integrable and, moreover, maximally superintegrable. We conclude in section~\ref{sec:last}. In the appendix A we provide explicit expressions for integrals of motion for $N=2$ and $N=3$.

\section{Description of the Model}
\label{sec:model}
\begin{figure*}[th!]
  \begin{center}
        \includegraphics[height=6cm]{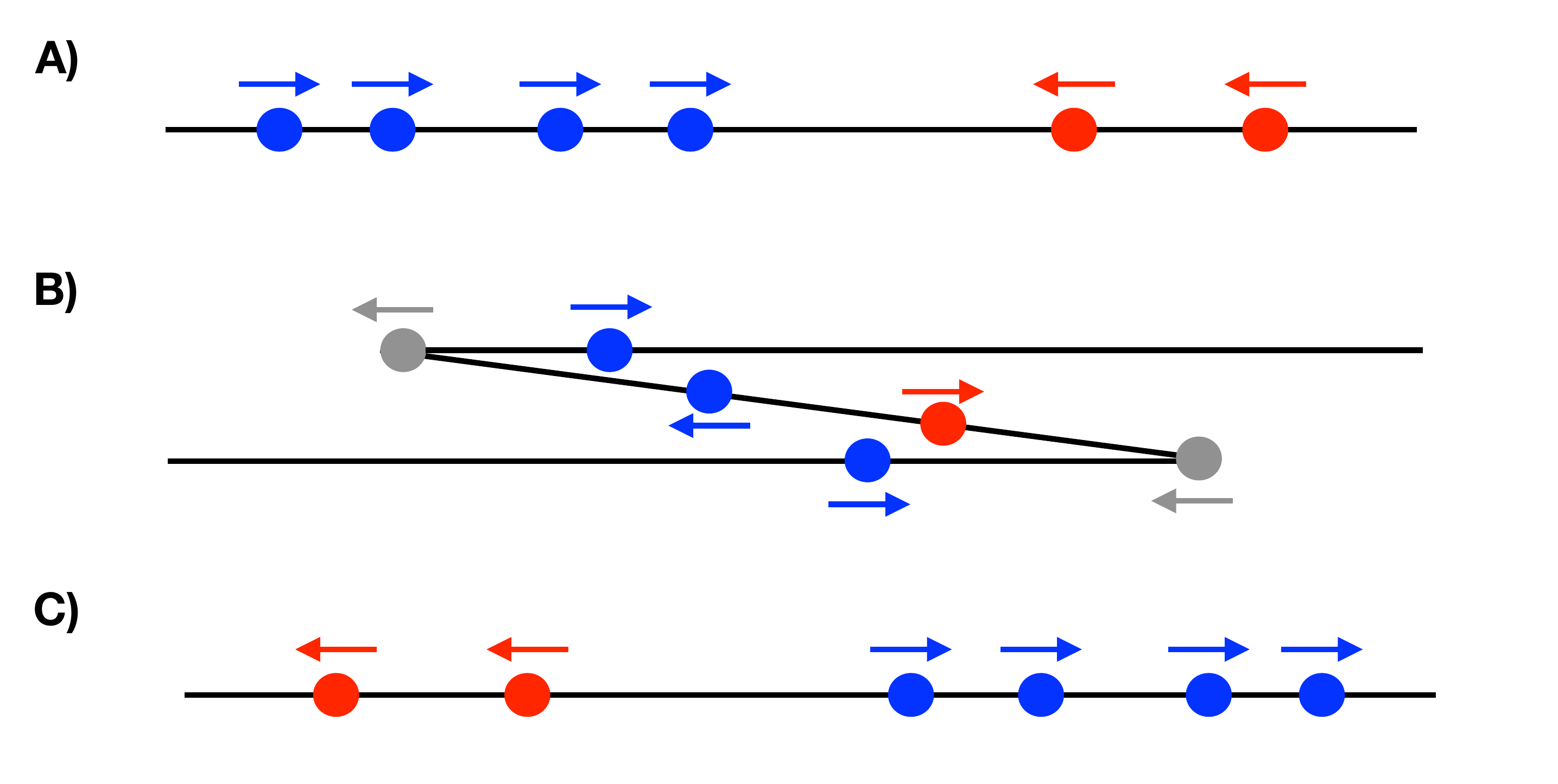} 
           \caption{Snapshots of  time evolution in the zigzag model. At early times A) all left-movers are located on the right and right-movers are on the left.
           The interaction period  B) proceeds through a series of zigzag formation resulting in the momenta exchanges. At late times C) all left-movers are on the left and right-movers are on the right. The topological invariant (\ref{Ttot}) ensures that the difference between the number of left- and right movers stays constant at all times, where left(right) is defined w.r.t. to the string worldsheet. This definition is illustrated by the color coding, where right-movers are colored blue and left-movers red.}
        \label{fig:T}
    \end{center}
\end{figure*}

The zigzag model describes a chain of $N$ ordered particles on a line with nearest neighbor interactions.
The structure of its Hamiltonian is very similar to the celebrated Toda chain \cite{Bazhanov_2017}.
The difference is that particles in the zigzag model are massless, {\it i.e.} they always move with unit velocity.
The exponential nearest neighbor Toda potential is replaced with a piecewise linear one of the form
\be
\label{pot}
{\cal V}(q)=q+|q|\;.
\ee
The full Hamiltonian then takes the following form
\be
\label{H}
H=\sum_{i=1}^N|p_i|+\sum_{i=1}^{N-1}{\cal V}(q_{i,i+1})\;,
\ee
where
\[
q_{i,i+1}=q_i-q_{i+1}\;.
\]
Pictorially, this system may be represented as a sequence of beads on a rubber band, see Fig.~\ref{fig:T}. All particles move with velocities $\pm1$ depending on the sign of the corresponding momentum. 
A generic configuration of particles exhibits a number of zigzags, which explains  the name of the model. The only particles experiencing a force are those at the zigzag turning points. Momenta of all other particles stay constant. 

In the gauge theory language, beads correspond to quarks in the adjoint representation and the rubber band to the confining string. As a consequence of asymptotic freedom, processes which change the number of quarks (partons) are suppressed at high energies. Hence the worldsheet theory splits into separate sectors labeled by the number of partons $N$, each described by (\ref{H}) at the leading order in the high-energy expansion.

Similarly to the Toda chain, in addition to the open zigzag model described by (\ref{H}), one may also consider its compact version. The latter describes a closed confining string wound around a compact spatial circle. As mentioned in   \cite{Donahue:2019adv}, there is  overwhelming numerical evidence that the compact zigzag model is also integrable. In the present paper we restrict to the open case.

An important property of the zigzag model is that it inherits Poincar\'e symmetry from the underlying gauge theory.
Namely, it is straightforward to check that the Poisson brackets between the Hamiltonian $H$, total momentum 
\be
\label{P}
P=\sum_{i=1}^Np_i\;,
\ee
and the boost generator 
\be
\label{J}
J=\sum_{i=1}^Nq_i|p_i|+{1\over 2}\sum_{i=1}^{N-1}(q_i+q_{i+1}){\cal V}(q_{i,i+1})
\ee
give rise to the $ISO(1,1)$ Poincar\'e algebra
\be
\label{HPJ}
\{ H,P\}=0\;,\;\;\{J,P\}= H\;,\;\;\{J, H\}=P\;.
\ee
As we will see, it is often convenient to treat momenta $p_i$ and coordinate differences $q_{i,i+1}$ on equal footing. This is achieved by introducing a string of variables
\be
Q_a=(p_1,q_{1,2},p_2,\dots,q_{N-1,N},p_N)\;
\ee
with $a=1,\dots, 2N-1$. Associated with this string there is also a sequence of the corresponding sign variables (``classical bits")
\be
S_a=(s_1,s_{1,2},s_2,\dots,s_{N-1,N},s_N)\;,
\ee
where 
\be
s_i=\mbox{ sign}(p_i)\;,\;\; s_{i,i+1}=\mbox{ sign}(q_{i,i+1})\;.
\ee
In what follows we also often use the notation
\be
S_0=S_{2N}=-1\;.
\ee
With these notations the equations of motion take the following simple form
\be
\label{eom}
\dot Q_a=S_{a-1}-S_{a+1}\;.
\ee

For any configuration of $Q_a$'s at early times one finds a bunch of right-moving particles on the left and a bunch of left-moving particles on the right, freely approaching each other ({\it i.e.}, no zigzags are present). 
As left- and right-movers reach each other and start to collide, the string goes through a sequence of zigzag configurations (see Fig.~\ref{fig:T}). At late times all zigzags are gone and one finds a bunch of left-movers on the left and a bunch of right-movers on the right. 

Classical integrability of the model manifests itself in the absence of  momentum exchange as one compares early and late configurations.  Namely, the values (and orderings) of all early and late left- and right-moving momenta are the same. Of course, particles momenta do change their values at intermediate times. The main goal of this paper is to prove integrability by constructing a sufficiently large set of conserved charges.

Note that any solution to (\ref{eom}) is a piecewise linear function of time. Hence, the zigzag model is defnitely integrable (or, better to say, solvable) in the broad sense---starting with any initial value it is straightforward to  find an explicit solution for a later time evolution. All that one needs to do is to linearly evolve the system forward in time untill one of the sign differences $S_{a-1}-S_{a+1}$ in the r.h.s. of (\ref{eom}) changes its value. This correspond either to a zigzag formation/annihilation ({\it i.e.} to a collision of two particles), or to a sign flip of one of the momenta. Then one continues linear evolution with different values of the velocities. A {\it Mathematica} solver implementing this procedure can be downloaded at  \url{https://jcdonahue.net/research}. Note that this solver provides an {\it exact} rather than a numerical solution to the equations of motion starting with arbitrary rational initial conditions. We will prove now that in addition to being integrable in this broad sense the zigzag model is actually Liouville integrable and, moreover, maximally superintegrable. 
\section{Topological Charge}
\label{sec:top}
Let us start a construction of conserved charges in the zigzag model by describing the topological charge introduced in \cite{Donahue:2019adv}. Namely, it is straightforward to check that 
\be
\label{Ttot}
T_{2N}={1\over 2}\sum_{a=0}^{2N-1} S_a S_{a+1}
\ee
stays constant under the time evolution described  by the equations of motion (\ref{eom}). We refer to $T_{2N}$ as topological charge because it defines a piecewise constant function on the phase space, separating it into dynamically disconnected topological sectors.
In the asymptotic regions $t\to\pm \infty$ no zigzags are present, hence all 
\[
s_{i,i+1}=-1
\]
so that one finds
\be
\left.T_{2N}\right|_{t\to\pm\infty}=-\sum_{i=1}^N s_i=N_L-N_R\;,
\ee
where $N_L$ and $N_R$ count the numbers of left- and right-movers in the initial and final states.  This proves that scattering does not change $N_L$ and $N_R$. To see the geometrical meaning of $T_{2N}$ at intermediate times  let us rewrite it in the following form
\be
\label{Ttot1}
T_{2N}={1\over 2}\sum_{i=1}^{N} s_i (s_{i-1,i}+s_{i,i+1})\;.
\ee
We see that also at intermediate times $T_{2N}$ can be interpreted as a difference in the number of left- and right-movers, provided left- and right-  is determined w.r.t. to the string worldsheet rather than w.r.t. to the physical space. Particles at the zigzag turning points should not be counted at all, see Fig.~\ref{fig:T}.
Interestingly, if one thinks about $S_a$'s as a classical spin sequence, $T_{2N}$ is equal to the Ising model Hamiltonian.

In the construction of the dynamical conserved charges presented in section~\ref{sec:linear} we will encounter the following piecewise constant functions on the phase space, which generalize (\ref{Ttot}),
\be
\label{Tab}
T_a={1\over 2}\sum_{b=0}^{a-1}S_bS_{b+1}\;.
\ee
Unlike $T_{2N}$, a general $T_{a}$ may change its value in the course of evolution when zigzags form/annihilate or particle momenta flip sign. Indeed using the equations of motion (\ref{eom}) we find that for $a<2N$
\be
\label{Taeom}
\dot{T}_a = \frac{1}{2} S_{a-1} \dot{S}_a = (1 - S_{a-1}S_{a+1} ) \delta(Q_a) \;.
\ee

\begin{figure*}[th!]
  \begin{center}
        \includegraphics[height=6cm]{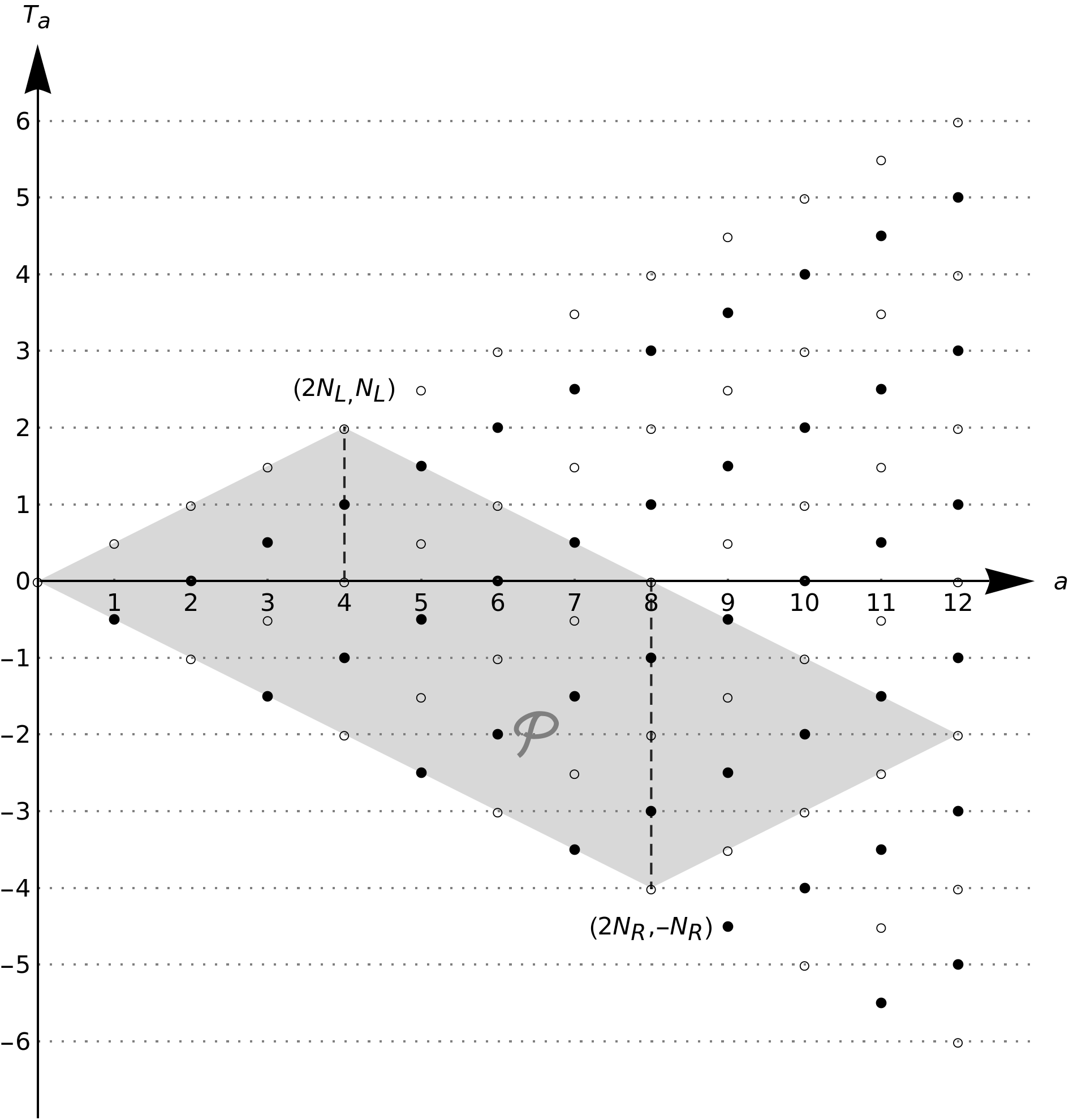} 
           \caption{ The shaded ${\cal P}$ region shows possible values of $T_a$ at fixed $N_L$, $N_R$. For solid dots  $S_a=1$, and for empty dots $S_a=-1$.}
        \label{fig:Tval}
    \end{center}
\end{figure*}

In what follows we need to know what are the possible values of 
\[
T_a\equiv T(0,a)
\]
 at fixed $N_L$, $N_R$ (or, equivalently, at fixed $N$, $T_{2N}$). In general, one can write
that
\be
\label{T0an}
T_a={a\over 2}-n_f\;,
\ee
where $n_f$ is the number of sign flips in the $(S_0,\dots,S_a)$ sequence of bits. In the absence of any additional restrictions, the possible range of values for $n_f$ is
\be
\label{nf}
0\leq n_f\leq a\;.
\ee
Restricting to the topological sector with a fixed $N_L$, $N_R$ imposes an additional constraint
\be
\label{Tcons}
n_f+\bar{n}_f=2N_R\;,
\ee
where 
\be
\label{barn}
0\leq \bar{n}_f\leq 2N-a\;
\ee
is a number of sign flips in the complementary sequence of bits $(S_a,\dots,S_{2N})$. Combining (\ref{Tcons}) and (\ref{barn}) we obtain that in addition to (\ref{nf}) the range of $n_f$ is also constrained to satisfy
\be
\label{nfLR}
a-2N_L\leq n_f\leq 2 N_R\;.
\ee
Recalling the relation (\ref{T0an}) between $n_f$ and $T_a$ the inequalities (\ref{nf}) and (\ref{nfLR}) imply that $T_a$ may take values in the shaded region
${\cal P}$ shown in Fig.~\ref{fig:Tval}. 

Fig.~\ref{fig:Tval} also illustrates another point, which will be important in Section~\ref{sec:linear}. Namely, as follows from  (\ref{T0an}), the value of $T_a$ unambiguously determines the value of the corresponding spin  $S_a$. This relation is shown in Fig.~\ref{fig:Tval}, where solid dots correspond to $S_a=1$ and empty ones to $S_a=-1$.

\section{Linear Charges}
\label{sec:linear}
Let us turn now to the construction of the dynamical conserved charges in the zigzag model. Given that the general solution of (\ref{eom}) is a piecewise linear function of time, it is natural to look for charges which are piecewise linear functions in the phase space. Restricting to translationally invariant charges we arrive then at the following ansatz,
\be
\label{linear}
I=\sum_{a=1}^{2N-1}F_a(S)Q_a\;.
\ee
A time derivative of $I$ contains a smooth contribution related to time evolution of $Q_a$'s and $\delta$-functional contributions caused by sign flips in the set of $S_a$'s. All $\delta$-functional contributions have to vanish separately, implying that $\dot{F}_a(S) \propto \delta(Q_a)$ or equivalently that the coefficient functions $F_a(S)$ satisfy 
\be
\label{Feqs}
\d_b F_a(S)(S_{b-1}-S_{b+1})=0\; \mbox{for } a\neq b\;.
\ee
It's clear by (\ref{Taeom}) that $T_a$ and $S_a$ satisfy this requirement. Using the equations of motion (\ref{eom}) it is straightforward to check that these are the only nontrivial solutions to (\ref{Feqs}) and that the coefficient functions take the following functional form
 \be
 \label{Fs}
 F_a(S)=F_a(S_a, T_a,N_L,N_R)\;.
 \ee
 In what follows we suppress the $N_L$ and $N_R$ dependence of $F_a$, assuming that these are kept fixed and non-zero. In addition, as discussed in Section~\ref{sec:top}, the value of $S_a$ is determined by $T_a$, so in what follows we write simply $F_a(T_a)$.

Then the single remaining equation comes from requiring that $I$ stays constant under a smooth time evolution of $Q^a$'s and takes the following form
\be
\label{Ieq}
C\equiv\sum_{a=1}^{2N-1}F_a(T_a)(S_{a-1}-S_{a+1})=0\;.
\ee
In particular, (\ref{Ieq}) implies that the value of $C$ is invariant under the change $S_a\to -S_a$, provided the flip of $S_a$ is dynamically allowed. By (\ref{eom}) a flip is allowed when 
\be
\label{flipcond}
S_{a-1}+S_{a+1}=0\;,
\ee
which gives the constraints
\be
C(S_a)|_{S_{a-1}=-S_{a+1}} =C(-S_a)|_{S_{a-1}=-S_{a+1}} \;.
\ee
These reduce to the following set of equations
\begin{widetext}
\be
\label{Feq}
F_{a+1}(T_{a-1})=S_{a-1}\l F_a\l T_{a-1}-{S_{a-1}\over 2}\r-F_a\l T_{a-1}+{S_{a-1}\over 2}\r\r+F_{a-1}(T_{a-1})\;.
\ee
\end{widetext}
These equations hold for any $a=1,\dots 2N-1$, if one sets $F_0=F_{2N}=0$.
 Roughly speaking, (\ref{Feq}) provide a set of linear recursion relations which determine $F_{a+1}$ in terms of $F_a$ and $F_{a-1}$. This is not exactly the case though, because $T_{a-1}$ (which appears 
 as an argument of $F_{a-1}$ in (\ref{Feq})) does not take all  values which $T_{a+1}$ may take, see Fig.~\ref{fig:Tval}. As a result  (\ref{Feq}) leaves 
 $F_{a+1}(T_{a+1})$ undetermined at
 \[
 T_{a+1}=\pm{a+1\over 2}\;.
 \]
 \begin{figure*}[th!]
  \begin{center}
        \includegraphics[height=6cm]{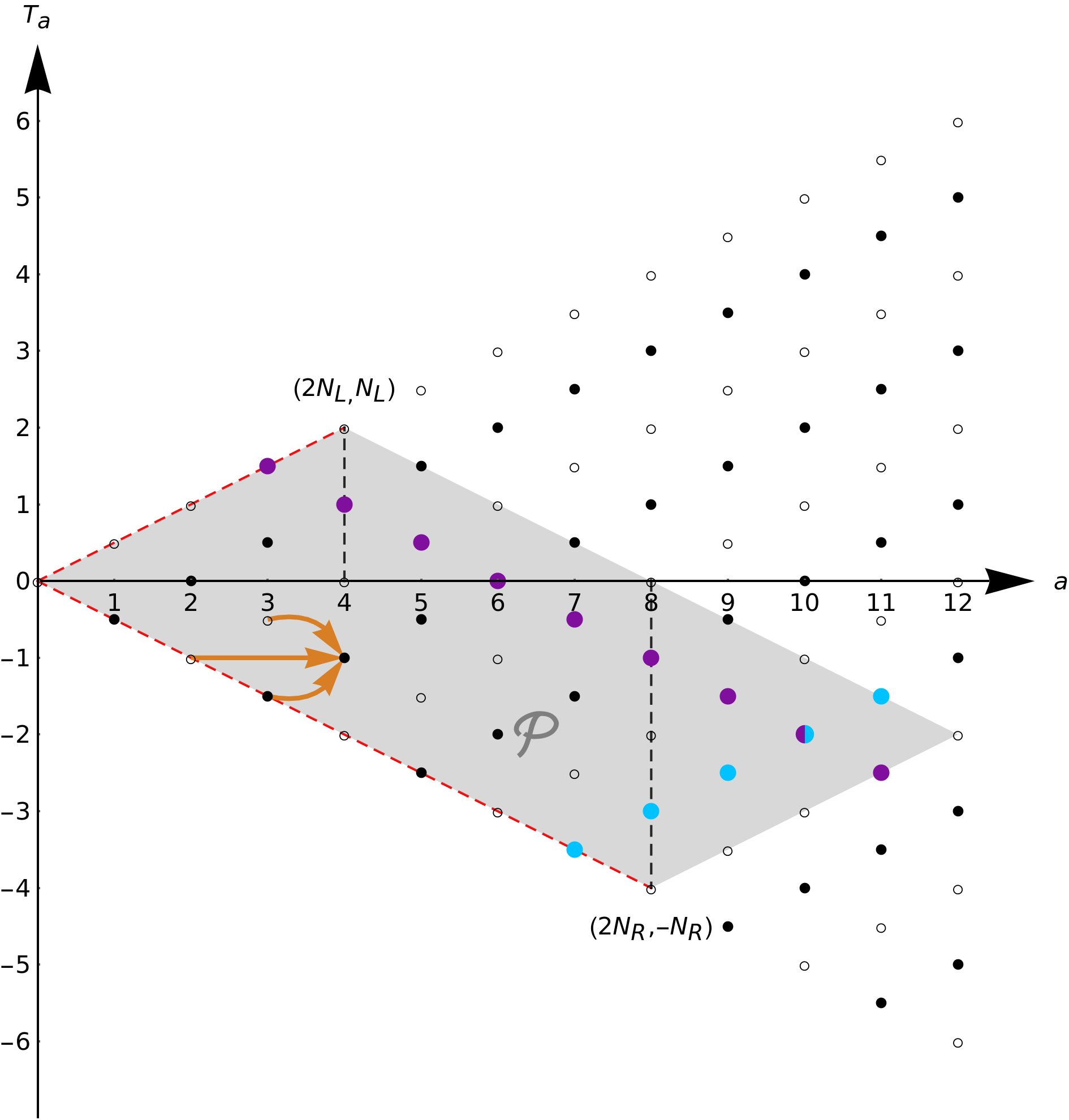} 
           \caption{ The structure of the recursion relation (\ref{Feq}) as illustrated by arrows in the $(a,T_a)$ plane. Blue and violet diagonals correspond to right and left charges respectively. Boundary conditions for the recursion relations (\ref{Feq}) are imposed at the points along the dashed red lines. }
        \label{fig:Tvalrec}
    \end{center}
\end{figure*}
 The structure of these recursion relations is illustrated in Fig.~\ref{fig:Tvalrec}, where we use the grid of possible values of $T_a$ from Fig.~\ref{fig:Tval} with the understanding that there is a number, which is the corresponding value of $F_a(T_a)$, assigned to each
 point of the grid. Arrows illustrate the relations between numbers at different points on the grid, as determined by (\ref{Feq}). Fig.~\ref{fig:Tvalrec} makes it clear that a general solution to (\ref{Feq}) is determined  by $2N$ boundary values  $F_a\l\pm {a+1\over 2}\r$, subject to one linear constraint at the right corner of the shaded rectangular ${\cal P}$ region, 
 \[
 F_{2N}=0\;.
 \]
This leaves us with $2N-1$ linearly independent solutions to the recursion relations (\ref{Feq}).
Not all of these solutions correspond to conserved charges, because (\ref{Feq}) is the condition for $C$ in (\ref{Ieq}) to be constant, rather than zero. Enforcing $C=0$ provides an additional linear constraint leaving us with 
$2N-2$ translationally invariant independent integrals of motion.
 
It is straightforward to construct these integrals explicitly.   Indeed, let us consider $F_a$'s which are non-zero only on one of the internal diagonals of the ${\cal P}$ region as shown in Fig.~\ref{fig:Tvalrec} and is equal to the corresponding $S_a$ at each of the points on that diagonal.  It is easy to see that these provide $2N-2$ linearly independent solutions to (\ref{Feq}). An explicit formula for the corresponding $F_a$'s is
\begin{gather}
F_a^{L,n_L}=S_a\delta_{T_a,n_L-{a\over 2}}
\end{gather} 
 for diagonals going from the upper left side of the ${\cal P}$ region to the lower right (like the violet one in Fig.~\ref{fig:Tvalrec})
 and
 \begin{gather}
F_a^{R,n_R}=S_a\delta_{T_a,-n_R+{a\over 2}}
\end{gather} 
for diagonals going from the lower left side to the upper right (like the blue one  in Fig.~\ref{fig:Tvalrec}). Here the range of values for $n_L,n_R$ is
\be
n_{L(R)}=1,\dots,2N_{L(R)}-1\;,
\ee
and $\delta_{T_a,n}$ is the Kronecker symbol. As a function of $S_a$'s the latter can be written as
\[
\delta_{T_a,n}=\prod_{ k=0}^{k\neq n+{a\over 2},k=a}{T_a+{a\over 2}-k\over n+{a\over 2}-k}\;,
\]
where $n$ can take any of the values $-{a\over 2},-{a\over 2}+1,\dots,{a\over 2}$. 

To see the physical meaning of these solutions let us inspect the corresponding functions $I^{L,n_L}$,  $I^{R,n_R}$ in the infinite past and future, $t\to \pm\infty$.
This is conveniently done by using the following interesting space-time interpretation of Fig.~\ref{fig:Tvalrec}. Note, that any particle configuration is naturally represented by a slice of ${\cal P}$.
Indeed, any configuration $Q_a(t)$ leads to a ``bit" sequence $S_a(t)$, which can be equivalently represented as a sequence of $T_a$ values, such that 
\[
T_{a+1}(t)=T_{a}(t)\pm {1\over 2}\;,
\]
\begin{figure*}[th!]
  \begin{center}
        \includegraphics[height=6cm]{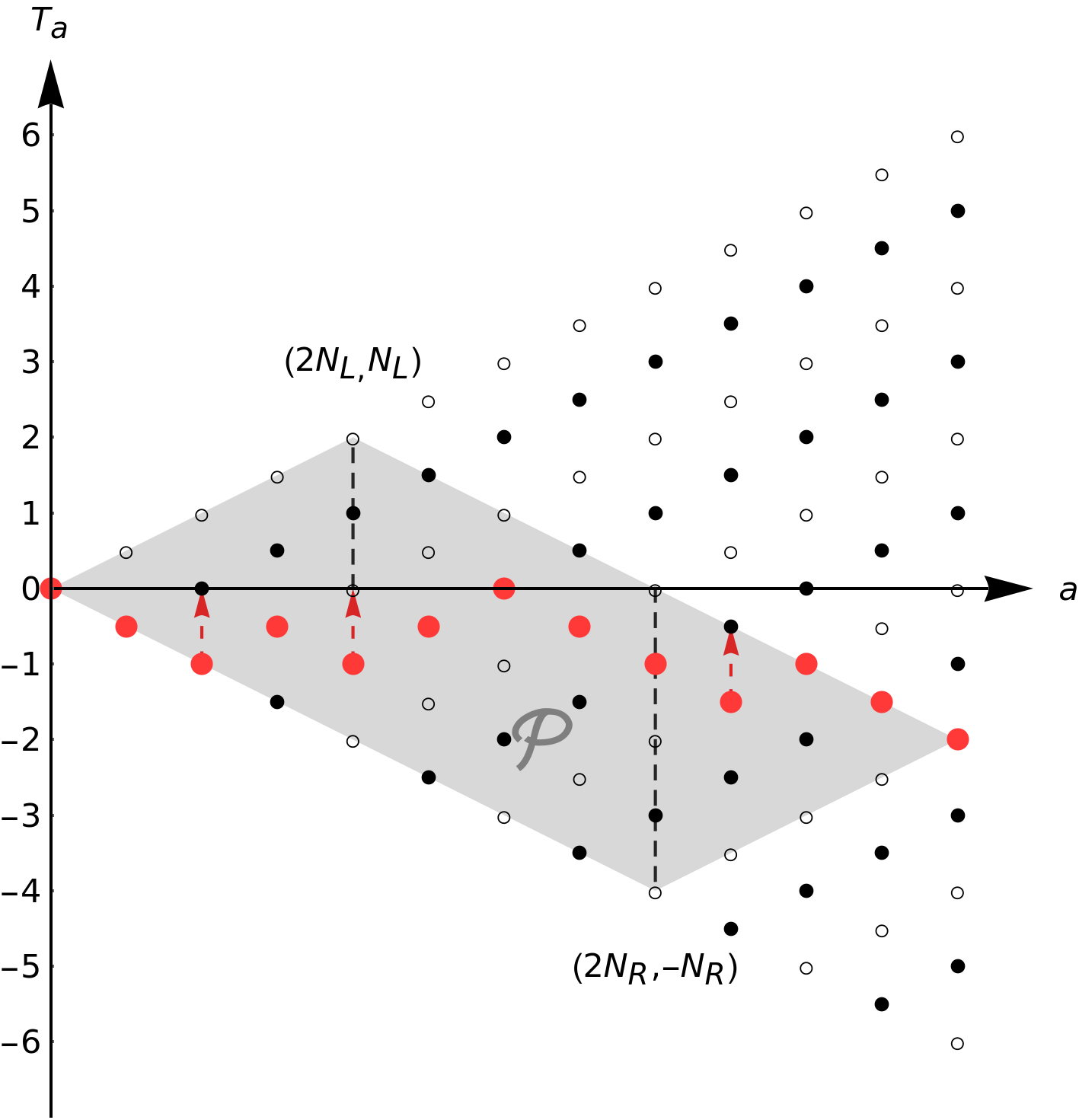} 
           \caption{A physical configuration of particles can be represented as a (red) slice in the $(a,T_a)$ plane. This snapshot corresponds to the one in Fig.~\ref{fig:T}B). Physical time evolution corresponds to the melting dynamics of this slice. Red dashed arrows show next possible 
           changes for the shape of the slice in the course of the time evolution.}
        \label{fig:Tvalslice}
    \end{center}
\end{figure*}
see Fig.~\ref{fig:Tvalslice}. At early times  no zigzags are present (i.e., all $q_{i,i+1}=-1$) and all left-movers are on the right and right-movers are on the left. Hence, this configuration corresponds to the values of $T_a$'s at the lower boundary of the ${\cal P}$ region. As time evolves the slice moves upwards monotonically. This motion corresponds to the dynamics of a melting 2D crystal---the evolution
proceeds through a series of upward jumps of the points at the corners of the ``melting surface".  At late times $t\to +\infty$ the slice reaches the upper boundary of ${\cal P}$.

Using this picture we see that at $t\to\pm \infty$
\begin{gather}
\label{Lasym} I^{L,n_L}=-Q^L_{n_L}\\
I^{R,n_R}=(-1)^{n_R+1}Q^R_{n_R}\;,
\end{gather}
where $Q^L_{n_L}$ and $Q^R_{n_R}$ are subsets of $Q_a$ corresponding to left- and right-movers at $t\to\pm \infty$, {\it i.e.}
at $t\to -\infty$
\begin{gather}
Q^R_{n_R}=Q_{n_R}
\\
Q^L_{n_L}=Q_{2N_R+n_L}
\end{gather}
and at $t\to +\infty$
\begin{gather}
Q^R_{n_R}=Q_{2N_L+n_R}
\\
\label{Qasym}  Q^L_{n_L}=Q_{n_L}
\;.
\end{gather}
Given that solutions of (\ref{Feq}) are either integrals of motion or linear functions of time, we see that $I^{L,n_L}$,  $I^{R,n_R}$  are actual integrals of motion (because they stay constant in the asymptotic regions). In particular, these integrals contain individual momenta of particles in the asymptotic regions $t\pm \infty$ so this construction proves that the set of initial and final momenta are conserved in the course of the collision (as well as the ordering of the momenta among left- and right-movers). It also proves the Liouville integrability of the system, because integrals corresponding to the asymptotic momenta provide us a set of $N$ commuting conserved charges.

The remaining $(N-2)$ constructed integrals in the asymptotic regions reduce to the  coordinate differences among left-movers or right-movers. Their existence implies that the time delays experienced by all left-movers are equal to each other and the same is true for the time delays experienced by all right-movers. 

To find these time delays let us inspect the last remaining independent solution of (\ref{Feq}). For  reasons which will become clear soon, we  refer to the corresponding piecewise linear function on the phase space (\ref{linear}) as
  $\tilde{H}$\footnote{This quantity is different from the one which was called $\tilde{H}$ in \cite{Donahue:2019adv}, but has similar properties.}. The corresponding $F_a$'s are non-zero at the right boundary of the ${\cal P}$ region. Unlike for internal diagonals, we need to use now both upper and lower parts of the boundary to satisfy the $F_{2N}=0$ condition. 
  This results in the following non-vanishing $F_a$'s for this solution 
\be
F^{\tilde{H}}_a=S_a\l \delta_{T_a,2N_L-{a\over 2}}-\delta_{T_a,-2N_R+{a\over 2}}\r\;.
\ee
Then in the asymptotic regions one finds
\be
\label{Htass}
\tilde{H}=
\left\{
\begin{array}{c}
Q_R-Q_L-P_L\;,\;\;\mbox{ at } t\to-\infty\\
Q_R-Q_L+P_R\;,\;\;\mbox{ at } t\to+\infty\;,
\end{array}
\right.
\ee
where $P_{L(R)}$ is the total asymptotic left(right)-moving momentum\footnote{It is defined in such a way that $P_{L(R)}\geq 0$.} and $Q_L$, $Q_R$ are the positions of the right-most left- and right-movers in the asymptotic regions. 
This implies that $\tilde H$ is a linear function of time, rather than a conserved charge, {\it i.e.}
\be
\label{Hevol}
\tilde{H}=2(t-t_0)\;,
\ee
where $t_0$ is a constant. Eqivalently, the Poisson brackets of $\tilde{H}$ with the Hamiltonian $H$ and momentum $P$ are
\be
\{ \tilde{H},H\}=2\;,\;\;\{P,\tilde{H}\}= 0\;,
\ee
where the latter follows from the translational invariance of $\tilde{H}$. This allows one to construct a new conserved charge
\be
\label{Pt}
\tilde{ P}=\{J,\tilde H\}\;,
\ee
which is not translationally invariant,
\be
\{ H,\tilde{P}\}=0\;,\;\;\{P,\tilde{P}\}= 2\;.
\ee
Altogether,  $I^{L,n_L}$,  $I^{R,n_R}$ and   ${\tilde P}$  provide a set of $(2N-1)$ independent conserved charges, which proves that the zigzag model is maximally superintegrable.

To calculate the time delays, let us evaluate $\tilde P$ in the asymptotic regions. Using the asymptotic expression (\ref{Htass}) we obtain
\be
\label{Ptass}
\tilde{P}=
\left\{
\begin{array}{c}
-Q_L-Q_R+P_L\;,\;\;\mbox{ at } t\to-\infty\\
-Q_L-Q_R+P_R\;,\;\;\mbox{ at } t\to+\infty\;,
\end{array}
\right.
\ee
Combining (\ref{Htass}), (\ref{Hevol}), (\ref{Ptass}) and the conservation of $\tilde{P}$ we find that
\be
\label{delays}
\Delta t_{L(R)}=P_{R(L)}
\ee
for the time delays $\Delta t_{L(R)}$ experienced by left(right)-moving particles. These time delays correspond to the celebrated shock wave phase shift \cite{'tHooft:1987rb,Amati:1987wq} confirming that the zigzag model describes the $N$-particle subsector of a massless $T\bar{T}$-deformed fermion.

\section{Discussion}
\label{sec:last}
To summarize, in this paper we presented an exhaustive analysis of the integrable structure of the classical zigzag model (\ref{H}). The natural next step is to quantize the model. Given that the classical time delay (\ref{delays}) reproduces the exact phase shift of a known quantum model---a massless $T\bar{T}$ deformed fermion---one expects the quantization preserving the Poincar\'e symmetry and integrability to exist. 

Note that a close relative of the zigzag model appeared in mid 70's under the name of  folded strings \cite{Bardeen:1975gx,Bardeen:1976yt}\footnote{We thank Antal Jevicki for directing us to these early papers.}. There,
exact solvability of a very similar Hamiltonian was understood as a consequence of the map between the corresponding mechanical solutions and folded string solutions of the two-dimensional Nambu--Goto theory. This correspondence reinforces the relation of the zigzag model with the $T\bar{T}$-deformation, given that the latter can be understood as arising from the coupling of an undeformed quantum field theory to two-dimensional strings \cite{Dubovsky:2017cnj,Dubovsky:2018bmo}.  In this language the zigzag model describes dynamics of a long string, while the early papers  \cite{Bardeen:1975gx,Bardeen:1976yt} studied the short string sector.
The relation of folded strings to $aQCD_2$ was conjectured in \cite{Bars:1993sq}. The analysis of \cite{Dubovsky:2018dlk,Donahue:2019adv} makes this relation precise, by demonstrating how the zigzag model arises as a leading high energy approximation to the worldsheet dynamics. 

The possibility of a consistent covariant quantization of folded strings remains somewhat controversial (see, e.g., \cite{Hanson:1976ey,Artru:1983gm}). We think that the connections to the $T{\bar T}$ deformation and  $aQCD_2$ strongly suggest that such a quantization is possible, and should in fact be one-loop exact (at least in the long string sector, corresponding to the zigzag model). Hopefully, a detailed understanding of the classical integrable structure achieved in the present paper will help to resolve this.

{\it Acknowledgements.}
We thank  Ofer Aharony, Misha Feigin, Antal Jevicki,  David Kutasov,  Grisha Korchemsky,  Conghuan Luo, Nikita Nekrasov, Sasha Penin and Slava Rychkov for useful discussions. 
This work is supported in part by the NSF award PHY-1915219 and by the BSF grant  2018068.

\appendix*
\section{Explicit Integrals for N=2,3}
For illustrative purposes we report here the explicit expressions for the integrals of motion for $N=2$ and $N=3$ with $N_L=1$. For $N=2$, asymptotics are defined by $s_1=1$, $s_2=\text{-}1$ for $t=-\infty$ and $s_1=\text{-}1$, $s_2=1$ for $t=+\infty$. Our translationally invariant integrals are
\be
\label{IR1N2}
\begin{split}
    I^{R,1}& =\frac{p_1}{2}(1+s_1)+\frac{q_{1,2}}{2}(1 + s_{1,2}) +\frac{p_2}{8}(3-s_1\\
    & +3 s_2 -s_1 s_2 +s_{1,2}+s_1 s_{1,2} +s_2 s_{1,2} + s_1 s_2 s_{1,2})
\end{split}
\ee
and
\be
\label{IL1N2}
\begin{split}
I^{L,1}& =\frac{p_1}{2}(-1+s_1)+\frac{q_{1,2}}{2}(1 + s_{1,2}) +\frac{p_2}{8}(-3-s_1 \\
& +3 s_2 +s_1 s_2 -s_{1,2}+s_1 s_{1,2} +s_2 s_{1,2} - s_1 s_2 s_{1,2}) \,\,\, .
\end{split}
\ee
As outlined in (\ref{Lasym}-\ref{Qasym}), asymptotically 
\be
\label{IR1AssN2}
I^{R,1}=
\left\{
\begin{array}{c}
p_1\;,\;\;\mbox{ at } t\to-\infty\\
p_2\;,\;\;\mbox{ at } t\to+\infty\;,
\end{array}
\right.
\ee
and
\be
\label{IL1AssN2}
I^{L,1}=
\left\{
\begin{array}{c}
-p_2\;,\;\;\mbox{ at } t\to-\infty\\
-p_1\;,\;\;\mbox{ at } t\to+\infty\;.
\end{array}
\right.
\ee
Further, we have
\be
\label{HtN2}
\tilde{H}=\frac{q_{1,2}}{2}(s_1 - s_1 s_{1,2}) +\frac{p_2}{4}(3-s_1 s_2 +s_{1,2} + s_1 s_2 s_{1,2})
\ee

\be
\label{PtN2}
\begin{split}
\tilde{P}& =\frac{q_1}{2}(-1+s_{1,2}) +\frac{p_2}{4}(-s_1+3 s_2 +s_1 s_{1,2}+ s_2 s_{1,2})\\ 
&-\frac{q_2}{2}(2-s_1 s_2+2 s_{1,2}+s_1 s_2 s_{1,2})
\end{split}
\ee
In accordance with (\ref{Htass}) and (\ref{Ptass}), asymptotically we find
\be
\label{HtassN2}
\tilde{H}=
\left\{
\begin{array}{c}
q_1-q_2-|p_2|\;,\;\;\mbox{ at } t\to-\infty\\
q_2-q_1+|p_2|\;,\;\;\mbox{ at } t\to+\infty\;,
\end{array}
\right.
\ee
and
\be
\label{PtassN2}
\tilde{P}=
\left\{
\begin{array}{c}
-q_1-q_2+|p_2|\;,\;\;\mbox{ at } t\to-\infty\\
-q_2-q_1+|p_2|\;,\;\;\mbox{ at } t\to+\infty\;.
\end{array}
\right.
\ee

For $N=3$, $N_L=1$ asymptotics are defined by $s_1=s_2=1$, $s_3= \text{-}1$ for $t=-\infty$ and $s_1=\text{-}1$, $s_2=s_3= 1$ for $t=+\infty$. Our translationally invariant integrals are

\be
\label{IR1N3}
\begin{split}
   I^{R,1} & = \frac{p_1}{2} ( s_1+1 ) +\frac{q_{1,2}}{2} ( s_{1,2}+1  ) +\frac{p_2}{8} ( s_2 s_1 s_{1,2}\\
   & +s_1 s_{1,2}+s_2 s_{1,2}+s_{1,2}-s_2 s_1-s_1+3 s_2+3 )\\
   &  +\frac{q_{2,3}}{8} ( s_1 s_{1,2}+s_1 s_{1,2} s_{2,3}-s_1 s_{2,3}+s_2 s_{1,2}\\
   & +s_2 s_{2,3}+s_2 s_{1,2} s_{2,3} +2 s_{2,3}-s_1+s_2+2 )  \\
   & +\frac{p_3}{32} ( -s_2 s_1 s_{1,2}-s_2 s_3 s_1 s_{1,2}+3 s_3 s_1 s_{1,2}+3 s_1 s_{1,2}\\
   & -s_2 s_1 s_{2,3}-s_2 s_3 s_1 s_{2,3}-s_3 s_1 s_{2,3} +s_2 s_1 s_{1,2} s_{2,3}\\
   & +s_2 s_3 s_1 s_{1,2} s_{2,3}+s_3 s_1 s_{1,2} s_{2,3}+s_1 s_{1,2} s_{2,3}\\
   & -s_1 s_{2,3}+3 s_2 s_{1,2}-s_3 s_{1,2}-s_{1,2}+3 s_2 s_{2,3}\\
   & +3 s_2 s_3 s_{2,3}+3 s_3 s_{2,3}+s_2 s_{1,2} s_{2,3}+s_2 s_3 s_{1,2} s_{2,3}\\
   & +s_3 s_{1,2} s_{2,3}+s_{1,2} s_{2,3}+3 s_{2,3}+s_2 s_1\\
   & +s_2 s_3 s_1-3 s_3 s_1-3 s_1+3 s_2 s_3 s_{1,2}+s_2+s_2 s_3\\
   & +5 s_3+5 ) \,\,\, , 
\end{split}
\ee
and
\be
\label{IR2N2}
\begin{split}
    I^{R,2} & = \frac{q_{1,2}}{4} ( s_1 s_{1,2}+s_{1,2}-s_1-1 ) +\frac{p_2}{8} ( -s_2 s_1 s_{1,2} \\
    & +s_1 s_{1,2}+s_2 s_{1,2}-s_{1,2}+s_2 s_1-s_1+3 s_2-3 )\\
    & +\frac{q_{2,3}}{8} ( -s_1 s_2 s_{1,2}-s_1 s_2 s_{2,3}+s_1 s_2 s_{1,2} s_{2,3}\\
    & -s_{1,2}+s_{1,2} s_{2,3}+3 s_{2,3}+s_1 s_2-3)\\
    & +\frac{p_3}{16} ( -s_2 s_1 s_{1,2}+s_2 s_3 s_1 s_{1,2}+s_3 s_1 s_{1,2}-s_1 s_{1,2}\\
    & -s_2 s_1 s_{2,3}+s_2 s_3 s_1 s_{2,3}-s_3 s_1 s_{2,3}+s_2 s_1 s_{1,2} s_{2,3}\\
    & -s_2 s_3 s_1 s_{1,2} s_{2,3}+s_3 s_1 s_{1,2} s_{2,3}-s_1 s_{1,2} s_{2,3}\\
    & +s_1 s_{2,3}-s_2 s_{1,2}+s_2 s_3 s_{1,2}+s_3 s_{1,2}-s_{1,2}-s_2 s_{2,3}\\
    & +s_2 s_3 s_{2,3}-s_3 s_{2,3}-s_2 s_{1,2} s_{2,3}+s_2 s_3 s_{1,2} s_{2,3}\\
    & -s_3 s_{1,2} s_{2,3}+s_{1,2} s_{2,3}+s_{2,3}+s_2 s_1-s_2 s_3 s_1\\
    & -s_3 s_1+s_1-s_2+s_2 s_3+5 s_3-5 ) \,\,\, .
\end{split}
\ee

Equations for $I^{R,3}$ and $I^{L,1}$ are similar. Correspondingly we find 
\be
\label{IR1AssN3}
I^{R,1}=
\left\{
\begin{array}{c}
p_1\;,\;\;\mbox{ at } t\to-\infty\\
p_2\;,\;\;\mbox{ at } t\to+\infty\;,
\end{array}
\right.
\ee
and
\be
\label{IR2AssN3}
I^{R,2}=
\left\{
\begin{array}{c}
-q_{1,2}\;,\;\;\mbox{ at } t\to-\infty\\
-q_{2,3}\;,\;\;\mbox{ at } t\to+\infty\;.
\end{array}
\right.
\ee
Further, we have
\be
\label{HtN3}
\begin{split}
    \tilde{H} & = \frac{q_{1,2}}{4}( - s_1 s_{1,2}+s_{1,2}+s_1-1 )  \\
    & +\frac{p_2}{8} ( s_2 s_1 s_{1,2}+s_1 s_{1,2}+s_2 s_{1,2}+s_{1,2}\\
    & -s_2 s_1-s_1+3 s_2 +3 )+ \frac{q_{2,3}}{16} (-3 s_2 s_1 s_{1,2}\\
    & -s_1 s_{1,2}-3 s_2 s_1 s_{2,3}+3 s_2 s_1 s_{1,2} s_{2,3}\\
    & +s_1 s_{1,2} s_{2,3}-s_1 s_{2,3}-s_2 s_{1,2}-3 s_{1,2}-s_2 s_{2,3}\\
    & +s_2 s_{1,2} s_{2,3}+3 s_{1,2} s_{2,3}+5 s_{2,3}+3 s_2 s_1+s_1+s_2-5 ) \\
    & +\frac{p_3}{32} ( s_2 s_1 s_{1,2}+3 s_2 s_3 s_1 s_{1,2}+s_3 s_1 s_{1,2}\\
    & -5 s_1 s_{1,2}+s_2 s_1 s_{2,3}+3 s_2 s_3 s_1 s_{2,3}+s_3 s_1 s_{2,3}\\
    & -s_2 s_1 s_{1,2} s_{2,3}-3 s_2 s_3 s_1 s_{1,2} s_{2,3}-s_3 s_1 s_{1,2} s_{2,3} \\
    & -3 s_1 s_{1,2} s_{2,3}+3 s_1 s_{2,3}-5 s_2 s_{1,2}+s_2 s_3 s_{1,2}\\
    & +3 s_3 s_{1,2}+s_{1,2}-5 s_2 s_{2,3}+s_2 s_3 s_{2,3}-5 s_3 s_{2,3}\\
    & -3 s_2 s_{1,2} s_{2,3}-s_2 s_3 s_{1,2} s_{2,3}-3 s_3 s_{1,2} s_{2,3}\\
    & -s_{1,2} s_{2,3}+s_{2,3}-s_2 s_1-3 s_2 s_3 s_1-s_3 s_1+5 s_1-3 s_2\\
    & -s_2 s_3+5 s_3+15 )
\end{split}
\ee

\be
\label{PtN3}
\begin{split}
    \tilde{P} & = \frac{q_1}{4} (-s_1 s_{1,2}+s_{1,2}+s_1-1 ) + \frac{p_2}{8} ( s_2 s_1 s_{1,2}+s_1 s_{1,2} \\
    & +s_2 s_{1,2}+s_{1,2}-s_2 s_1-s_1+3 s_2+3 )  \\
    & + \frac{q_2}{16} (-s_2 s_1 s_{1,2} +5 s_1 s_{1,2}-s_2 s_1 s_{2,3}+s_2 s_1 s_{1,2} s_{2,3}\\
    & -s_1 s_{1,2} s_{2,3}+s_1 s_{2,3} +s_2 s_{1,2}-5 s_{1,2}\\
    & +s_2 s_{2,3}-s_2 s_{1,2} s_{2,3}+s_{1,2} s_{2,3}+7 s_{2,3}+s_2 s_1\\
    & -5 s_1-s_2-3 )+ \frac{q_3}{16} (-3 s_2 s_3 s_1 s_{1,2}-s_3 s_1 s_{1,2} \\
    & +4 s_1 s_{1,2}-3 s_2 s_3 s_1 s_{2,3}-s_3 s_1 s_{2,3}+3 s_2 s_3 s_1 s_{1,2} s_{2,3}\\
    & +s_3 s_1 s_{1,2} s_{2,3}+4 s_1 s_{1,2} s_{2,3}-4 s_1 s_{2,3}+4 s_2 s_{1,2}\\
    & -s_2 s_3 s_{1,2}-3 s_3 s_{1,2}+4 s_2 s_{2,3}-s_2 s_3 s_{2,3}+5 s_3 s_{2,3}\\
    & +4 s_2 s_{1,2} s_{2,3}+s_2 s_3 s_{1,2} s_{2,3}+3 s_3 s_{1,2} s_{2,3}-8 s_{2,3}\\
    & +3 s_2 s_3 s_1+s_3 s_1-4 s_1+4 s_2+s_2 s_3-5 s_3-8 ) \\
    & +\frac{p_3}{32} ( 3 s_2 s_1 s_{1,2}+s_2 s_3 s_1 s_{1,2}-5 s_3 s_1 s_{1,2}\\
    & +s_1 s_{1,2}+3 s_2 s_1 s_{2,3}+s_2 s_3 s_1 s_{2,3}+3 s_3 s_1 s_{2,3}\\
    & -3 s_2 s_1 s_{1,2} s_{2,3}-s_2 s_3 s_1 s_{1,2} s_{2,3}-3 s_3 s_1 s_{1,2} s_{2,3}\\
    & -s_1 s_{1,2} s_{2,3}+s_1 s_{2,3}+s_2 s_{1,2}-5 s_2 s_3 s_{1,2}+s_3 s_{1,2}\\
    & +3 s_{1,2}+s_2 s_{2,3}-5 s_2 s_3 s_{2,3}+s_3 s_{2,3}-s_2 s_{1,2} s_{2,3}\\
    & -3 s_2 s_3 s_{1,2} s_{2,3}-s_3 s_{1,2} s_{2,3}-3 s_{1,2} s_{2,3}-5 s_{2,3}-3 s_2 s_1\\
    & -s_2 s_3 s_1+5 s_3 s_1-s_1-s_2-3 s_2 s_3+15 s_3+5 )
\end{split}
\ee

In accordance with (\ref{Htass}) and (\ref{Ptass}), asymptotically we find
\be
\label{HtassN3}
\tilde{H}=
\left\{
\begin{array}{c}
q_2-q_3-|p_3|\;,\;\;\mbox{ at } t\to-\infty\\
q_3-q_1+|p_2|+|p_3|\;,\;\;\mbox{ at } t\to+\infty\;,
\end{array}
\right.
\ee
and
\be
\label{PtassN3}
\tilde{P}=
\left\{
\begin{array}{c}
-q_2-q_3+|p_3|\;,\;\;\mbox{ at } t\to-\infty\\
-q_3-q_1+|p_2|+|p_3|\;,\;\;\mbox{ at } t\to+\infty\;.
\end{array}
\right.
\ee
We see that these expressions rapidly become quite lengthy, in spite of the existence of a simple geometric description provided in the main text of the paper.

\bibliographystyle{utphys}
\bibliography{dlrrefs}
\end{document}